\documentclass[aps,prd,amsfonts,nofootinbib,supercriptaddress]{revtex4}

\usepackage{amssymb,amsmath}
\usepackage{graphicx}
\usepackage{epsfig}
\usepackage{color}

\newcommand{\beq}{\begin{equation}}
\newcommand{\eeq}{\end{equation}}
\newcommand{\bea}{\begin{eqnarray}}
\newcommand{\eea}{\end{eqnarray}}
\newcommand{\vc}[1]{{\textbf{#1}}}

\newcommand{\dd}{{\rm d}}
\newcommand{\fett}[1]{\boldsymbol{#1}}
\newcommand{\nabq}{\fett{\nabla}_{\fett{q}}}
\newcommand{\nabx}{\fett{\nabla}_{\fett{x}}}

\begin{document}

\title{A Relativistic view on large scale N-body simulations}

\author{Cornelius Rampf$\,^a$}

\author{Gerasimos Rigopoulos$\,^b$}

\author{Wessel Valkenburg$\,^c$}

\affiliation{\vspace{0.15cm}
$^a$ Institute of Cosmology and Gravitation, University of Portsmouth, Dennis Sciama Building, Burnaby Road, Portsmouth, PO1 3FX, United Kingdom \vspace{0.1cm}\\ $^b$ Institut f\"ur Theoretische Physik, Philosophenweg 12,
Universit\"at Heidelberg, 69120 Heidelberg, Germany\vspace{0.1cm}\\
$^c$ Instituut-Lorentz for Theoretical Physics, Universiteit Leiden,
       Niels Bohrweg 2, Leiden, NL-2333 CA, The Netherlands}

\begin{abstract}
\noindent
We discuss the relation between the output of Newtonian N-body simulations on scales that approach or exceed the particle horizon to the description of General Relativity. At leading order, the Zeldovich approximation is correct on large scales, coinciding with the General Relativistic result. At second order in the initial metric potential, the trajectories of particles deviate from the second order Newtonian result and hence the validity of 2LPT initial conditions should be reassessed when used in very large simulations. We also advocate using the expression for the synchronous gauge density as a well behaved measure of density fluctuations on such scales.

\end{abstract}

\maketitle

\section{Introduction}
The process of large scale structure formation has been studied using either analytical perturbation theory (Newtonian or Relativistic) \cite{Bernardeau:2001qr, Malik:2008im} or numerical N-body simulations. Traditionally, these two techniques were applicable on largely non-overlapping domains. Perturbation theory is valid when density contrasts and other perturbations are small, corresponding to early times and/or large cosmological scales. Its analytical nature is of course a great advantage but it quickly breaks down when non-linearities start to form. It also allows for an in-principle well understood transition from relativistic to non-relativistic theory, although work is still ongoing for better understanding this relation beyond the leading order
\cite{Green:2011wc, Hwang:2012ra, Flender:2012nq, Kopp:2013tqa, Villa:2014aja}. On the other hand, N-body simulations treat the full non-linear evolution by solving the newtonian equations for a collection of gravitationally interacting dark matter particles. Computational resources in general restrict the size of the simulated volumes but this is not necessarily a restriction of N-body simulations: all non-linear effects in our universe take place on relatively small scales compared to the hubble radius and it is these effects that N-body codes have traditionally targeted. Therefore analytic pertubative approaches and numerical N-body simulations have for the most part been studying two largely non-overlapping regimes of cosmological structure formation.

Recent trends and advances in cosmology have brought the quasi linear regime of structure formation into the focus of investigations, with the Baryon Acoustic Oscillation pattern (BAO) as a prominent feature \cite{Basset Hlozek}. Determining the power spectrum at the BAO regime ($k\sim 0.02 - 0.2$ h Mpc$^{-1}$) at (sub-)percent accuracy is crucial for determining the equation of state of dark energy from future cosmological surveys such as EUCLID \cite{Amendola:2012ys}. However, the BAOs lie at the borders of validity of standard analytic pertubative techniques and so in recent years significant effort has been devoted to extending analytic understanding into the quasi-linear regime - see \cite{Bernardeau:2013oda} and references therein. On the other hand, numerical simulations can provide a suitably accurate prediction of the BAO pattern but only if very large simulation volumes are employed, required to improve the statistics and bring uncertainties down to cosmic variance levels. The largest simulation to date, DEUS-FUR, has produced cosmic variance limited power spectra and required a simulation volume equal to the current particle horizon \cite{Rasera:2013xfa}.

Cosmological N-body simulations use Newtonian equations of motion to compute the trajectories of a set of particles interacting through Newtonian gravitational forces, with the Newtonian potential updated instantaneously via the Poisson equation throughout the simulation volume. However, for very large simulations, as the one mentioned above, the large scale peculiar motions of particles are determined by perturbations with wavelengths larger than than the particle horizon i.e. take place outside of causal contact. For $\Lambda CDM$ the comoving particle horizon is given by (ignoring radiation)
\beq
d_{\rm cp}(z)=\frac{1}{H_0}\int\limits_z^\infty\frac{du}{\sqrt{\Omega_\Lambda+\Omega_{\rm M}(1+u)^3}}=\frac{2}{H_0\sqrt{\Omega_M}}\frac{_2F_1\left(\frac{1}{6},\frac{1}{2},\frac{7}{6},-\frac{\Omega_\Lambda}{\Omega_M (1+z)^3}\right)}{\sqrt{1+z}}\,,
\eeq
where $_2F_1$ is the hypergeometric function. In fact, for large enough simulations there will be length scales lying outside of causal contact at early times, when initial conditions are set up, even if the simulation boxes are within the horizon at $z=0$, see eg fig 1. For example, the Millenium XXL simulation with box size of $3h^{-1}$ GPc lies outside the particle horizon before $z\simeq 50$. Even for moderately large simulations, initial conditions and the initial stages of evolution involve regions that are not in causal contact. Is it then correct to use Newtonian dynamics in large simulations or at the redshifts when initial conditions are set up? One possible justification is that particles move very little on super-horizon scales compared to the displacements during the non-linear evolution, which the N-body simulations tackle correctly. This is a reasonable argument but a more complete justification for the use of Newtonian dynamics and Newtonian gravity on such scales would be desirable, especially given the accuracy that large simulations aim for.

\begin{figure*}
\includegraphics[width=0.45\textwidth]{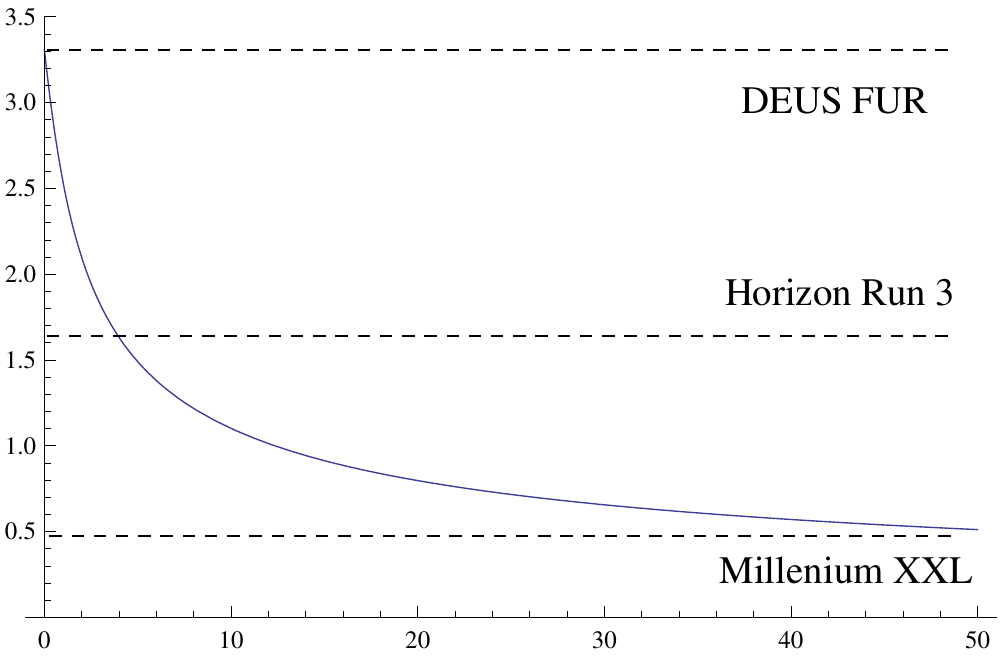}
\caption{The comoving particle horizon of $\Lambda CDM$ (solid line) compared with the comoving box size of recent large N-body simulations. The vertical axis denotes comoving distance in units of $1/H_0$ and the horizontal axis is redshift.}
\end{figure*}

In this short paper we address the question of the compatibility of Newtonian N-Body simulations with General Relativity when the particle horizon lies within the simulation box. The contents are largely based on \cite{Rigopoulos:2012xj, Rampf:2012pu, Rampf:2013ewa, Rigopoulos:2013nda, Rampf:2013dxa} to which we refer for more details. The strategy we adopt is the following: Since we are interested in the evolution on long wavelengths we first present the solution to the Einstein equations in a gradient expansion \cite{Comer:1994np, Stewart:1994wq, Rigopoulos:2012xj}. We do this in the synchronous comoving gauge which is akin to a lagrangian point of view where fluid elements are labeled by their initial spatial positions $\vc{q}$. The Eulerian coordinates (spatial and temporal) of the N-body simulation are identified with the Poisson gauge and the Lagrangian-to-Eulerian transformation in the relativistic setting is in fact determined by the gauge transformation between the synchronous-comoving and the Poisson gauge. We find that at leading order in the initial metric perturbation the Zeldovich approximation is recovered but differences emerge at second order. As we discuss, the second order relativistic effects in the trajectory of particles can easily be incorporated in standard Newtonian simulations. It is worth noting that since many N-body simulations use second order Lagrangian Perturbation Theory (2LPT) to set up initial conditions, our results imply that the accuracy usually claimed by using 2LPT formulae should be reexamined for large simulations. Although the effects are small, the investigation of this issue is worth pursuing given the accuracy required by upcoming dark energy studies and which current large N-body simulations target. We stress that we are concerned with large scales here. Simulations that can track relativistic effects on subhorizon scales are described in \cite{Adamek:2014xba} of this special issue.

\section{The synchronous metric for perturbed $\Lambda CDM$}
The first step is to derive the metric in the comoving-synchronous gauge with line element
\beq\label{co-synch}
ds^2=-\dd t^2+\gamma_{ij}(t,\vc{q}) \, \dd q^{i} \dd q^{j}\,,
\eeq
a gauge convenient for studying CDM on large scales. As we are interested in large scales, we use a gradient expansion to solve the Einstein Equations for $\gamma_{ij}$. Further restricting the gradient expansion to second order in the potential, the spatial metric takes the form
\begin{align}
  \gamma_{ij} (t,\fett{q}) = \,\,&a^2(t) \Bigg\{ \, \delta_{ij} \left( 1+ \frac{10}{3} \Phi \right)
    + 3\,D(t) \left[ \Phi_{,ij} \left( 1-\frac{10}{3} \Phi \right)
      - 5 \Phi_{,i} \Phi_{,j} +  \frac 56 \delta_{ij}  \Phi_{,l} \Phi_{,l} \right] \nonumber \\
   &\quad  + \frac{9}{4}  E(t)
   \Bigg[ 4\Phi_{,ll} \Phi_{,ij}   \Bigg. \nonumber
  -  \delta_{ij} \left( \Phi_{,ll} \Phi_{,mm} - \Phi_{,lm}\Phi_{,lm}  \right)   \Bigg] \nonumber \\
  &\quad +  \frac{9}{4}  \left[ D^2(t) -4 E(t) \right] \,\Phi_{,li} \Phi_{,lj} + {\cal O}(\Phi^3) \Bigg. \Bigg\} \,, \label{metric-phi}
\end{align}
where we have denoted differentiation w.r.t. the comoving coordinates $q^i$ with a comma and a summation over repeated indices is implied. Defining the auxiliary functions
\begin{align}
\label{evoJK}
  J(a) &= \frac{a^{-1}}{2H_0}\int^{a} \frac{\dd u}{\sqrt{\Omega_\Lambda+\Omega_Mu^{-3}}} \,,  \\
  K(a) &= \frac{a}{H_0}\int^{a} \frac{J^2(u)}{u^2\sqrt{\Omega_\Lambda+\Omega_Mu^{-3}}} \, \dd u  \,,
\end{align}
the time dependent coefficients in the metric are given by
\begin{align}
   D(a) &= \frac{1}{H_0}\frac{20}{9} \int^a \frac{J(u)}{u^3\sqrt{\Omega_\Lambda+\Omega_Mu^{-3}}} \, \dd u \,, \label{D(a)}\\
   E(a)  &=  \frac{1}{H_0}\frac{200}{81} \int^a \frac{1}{u^3\sqrt{\Omega_\Lambda+\Omega_Mu^{-3}}}   \left[ \frac{K(u)}{u^2}  -  \frac{9}{10} D(u) J(u)   \right]\dd u \,,
\end{align}
The relation of $\Phi$ to the usual curvature perturbation $\zeta$ is
\beq
1+\frac{10}{3}\Phi(\vc{q})={\exp}(2\zeta(\vc{q}))\quad\Rightarrow\quad \Phi (\vc{q}) = \frac{3}{5}\zeta(\vc{q})+\frac{3}{5}\zeta^2(\vc{q})+\ldots
\eeq

The above expressions contain enough arbitrary constants, reflected in the indefinite integrals, to set $D(a)$ and $E(a)$ to zero at any chosen initial time. From now on and for simplicity we will take $a\rightarrow 0$ as the initial time and assume all functions to be in their growing mode. It is worth noting that the gradient expansion, when restricted to second order in $\Phi$ coincides with a second order perturbative calculation that is performed without reference to long scales, see eg \cite{Russ:1995eu, Matarrese:1997ay}. Thus, for the case of CDM there are no short scale phenomena that are missed by a gradient expansion, at least until shell crossing which invalidates the comoving synchronous coordinate system. This is an interesting property of inhomogenous CDM, probably related to CDM lacking a sound horizon.

The integral defining the auxiliary function $J$  can be computed analytically
\beq
J(a)=\frac{a^{3/2}}{5H_0\sqrt{\Omega_M}} \,\,{}_2F_1\left(\frac{1}{2},\frac{5}{6},\frac{11}{6},-\frac{\Omega_\Lambda}{\Omega_M}a^3\right)
\eeq
where ${}_2F_1$ is the hypergeometric function. It is then possible to integrate (\ref{D(a)}) to obtain for $D(a)$
\beq
D(a)=\frac{1}{H_0^2\Omega_M}a\left(1+\frac{\Omega_\Lambda}{\Omega_M}a^3\right)^{1/2}
{}_2F_1\left(\frac{3}{2},\frac{5}{6},\frac{11}{6},-\frac{\Omega_\Lambda}{\Omega_M}a^3\right)\,.
\eeq
This is simply the $\Lambda CDM$ growth factor so $D(a)=D_+(a)$. The CDM density is given by
\beq\label{density}
\rho(t,\vc{q})=\frac{3H_0^2\Omega_{M}}{8\pi G}\frac{C(\vc{q})}{\sqrt{{\rm Det}\gamma_{ij}(t,\vc{q})}}\,,
\eeq
where $C(\vc{q})$ a time independent constant function of $\vc{q}$. Setting
\beq
C(\vc{q})=\left(1+\frac{10}{3}\Phi(\vc{q})\right)^{3/2}\,,
\eeq
and expanding to second order we find
\begin{widetext}
\begin{align}\label{density2}
\rho(t,\vc{q})=\frac{3H_0^2\Omega_{M}}{8\pi G}\frac{1}{a^3}\left(1-\frac{3}{2}D\Phi_{,ll}+10D\Phi\Phi_{,ll }+\frac{15}{4}D\Phi_{,l} \Phi_{,l}+\frac{9}{8}\left(D^2-E\right)\Phi_{,ll}\Phi_{,mm} +\frac{9}{8}\left(D^2+E\right)\Phi_{,lm}\Phi_{,lm}\right)\,,
\end{align}
\end{widetext}
which at linear order gives the growing mode of density perturbations in the comoving gauge.

\section{Transformation to the Poisson Gauge}
The spatial coordinate system used above is comoving with the CDM fluid, i.e., each particle is characterized by a fixed $\vc{q}$ throughout the evolution. All information about inter-particle distances and clustering is encoded in the metric. This however is not the most convenient way to compare with the output of an N-body simulation where a coordinate grid is fixed and particles move through the grid. Let us therefore define a coordinate transformation from the
comoving coordinates $(t,\vc{q})$ to another coordinate system $(\tau, \vc{x})$, the Poisson gauge, where we require the metric to take the form
\begin{align}
g_{00}(\tau,\vc{x})&=-\left[ 1+2A(\tau,\vc{x}) \right] \,, \\
g_{0i}(\tau,\vc{x})&=a(\tau)w_i \,, \\
g_{ij}(\tau,\vc{x})&=\delta_{ij}\left[ 1-2B(\tau,\vc{x}) \right ] a^2(\tau) \,,
\end{align}
where $A\ll 1$ and $B\ll 1$ and $w_i$ is a divergence-less vector ($\frac{\partial w_i}{\partial x^i}=0$). The coordinate transformation is written as
\begin{align}
x^i(t,\vc{q})  &= q^{i}+{F}^i(t,\vc{q})\label{x-xfn} \,, \\
\tau (t,\vc{q})&= t+{L}(t,\vc{q})\,,
\end{align}
and the metrics are then related through
\begin{align}
\gamma_{ij}(t,\vc{q}) &=-\frac{\partial \tau}{\partial q^{i}}\frac{\partial \tau}{\partial q^{j}} \left(1+2A(\tau,\vc{x})\right)+2\frac{\partial \tau}{\partial q^{i}}\frac{\partial x^{l}}{\partial q^{j}}a(\tau)w_l(\tau,\vc{x})+\frac{\partial x^{l}}{\partial q^{i}}\frac{\partial x^{m}}{\partial q^{j}}\delta_{lm}\left(1-2B(\tau,\vc{x})\right)a(\tau)^2 \,,\label{glm}\\
0&=-\frac{\partial \tau}{\partial t}\frac{\partial \tau}{\partial q^{i}}\left(1+2A(\tau,\vc{x})\right)+\left(\frac{\partial \tau}{\partial t}\frac{\partial x^{l}}{\partial q^{i}}+\frac{\partial x^{l}}{\partial t}\frac{\partial \tau}{\partial q^{i}}\right)a(\tau)w_l(\tau,\vc{x})+\frac{\partial x^{l}}{\partial t}\frac{\partial x^{m}}{\partial q^{i}}\delta_{lm}\left(1-2B(\tau,\vc{x})\right)a(\tau)^2\,,\label{g0i}\\
-1 &=-\left(\frac{\partial \tau}{\partial t}\right)^2\!\!\!\left(1+2A(\tau,\vc{x})\right)+2\frac{\partial \tau}{\partial t}\frac{\partial x^{l}}{\partial t}a(\tau)w_l(\tau,\vc{x})+\frac{\partial x^{l}}{\partial t}\frac{\partial x^{m}}{\partial t}\delta_{lm}\left(1-2B(\tau,\vc{x})\right)a(\tau)^2\,.\label{g00}
\end{align}
which can be solved perturbatively for the $L$ and $F^i$. We find

\begin{align}
 \tau&= t + v \,\Phi +   \frac 9 4  a^2\dot E \frac{1}{\nabq^2} \mu_2 +  \frac 9 4\frac{v D}{3} \Phi_{,l}\Phi_{,l}
       + v  \left( v H +\frac 3 4 a^2 \ddot D -\frac 5 3 \right) \Phi^2  +v  \left( 2 vH +3a^2 \ddot D + \frac{10}{3}\right) V , \label{t-xfn}\\
 x^i &=q^i + \Bigg(\frac 3 2 D \Phi  + \frac 9 4 E
    \frac{1}{\fett{\nabla}_{\fett{q}}^2} \mu_2 -5 D  \Phi^2 + \left( 5 D + \frac{v^2}{a^2} \right)   V \Bigg)_{,i}+ \left( 5 D + \frac{v^2}{a^2} \right) W_i \label{x-xfn2}\,,
\end{align}
where a comma denotes a partial derivative wrt the $q^i$ coordinates and we have defined
\begin{widetext}
\begin{align}
  v &\equiv \frac 3 2 a^2 \dot D \,,  \quad
  \mu_2 (\fett{q}) \equiv  \frac 1 2 \left( \Phi_{,ll} \Phi_{,mm} - \Phi_{,lm} \Phi_{,lm}  \right) \,, \\
\label{kernelC}  V  (\fett{q}) &\equiv \frac{3}{2}\frac{1}{\nabq^2 \nabq^2}  \mu_2 +\frac{1}{2}\frac{1}{\nabq^2}\left(\Phi_{,m} \Phi_{,m}\right)\,,\\
 \label{kernelR} W_i (\fett{q}) &\equiv \frac{1}{\nabq^2}  \left[  \Phi_{,i}\Phi_{,mm}  - \Phi_{,li} \Phi_{,l}
      -2\frac{\partial_i}{\nabq^2} \mu_2 \right]  \,,
 \end{align}
\end{widetext}
where $W_i$ is divergence-less vector. For the metric functions of the Poisson Gauge we have
\begin{align}
\label{A}
A(\tau,\vc{x})= \,\,&A_N  +\frac{25}{9}\left(\frac{3}{2}-\frac{18}{5}HJ+\frac{2}{5}J^2\Lambda\right)\Phi^2
-\frac{200}{9}\left(\frac{1}{2}-3HJ+7H^2J^2-J^2\Lambda\right) V \,,
\end{align}
\begin{align}
B(\tau,\vc{x}) = \,\,& B_N +\frac{25}{9}\left(\frac{2}{5}HJ-8H^2J^2+\frac{2}{5}J^2\Lambda\right)\Phi^2 -\frac{200}{9}HJ\left(2HJ-1\right)V\,,
\label{B}\\
\vc{w}=&-a(\tau)\frac{\partial}{\partial\tau}\left( 5 D + \frac{v^2}{a^2} \right) W_i\,,\label{w}
\end{align}
where we have defined
\begin{align}
A_N(\tau,\vc{x})=B_N(\tau,\vc{x})\equiv \frac{5}{3}\left( 2HJ-1\right)\Phi + \frac{5}{4}D\left[\Phi_{|l}\Phi_{|l}
 -2\left(2HJ-1-\frac{20}{9}\frac{\dot K}{D}\right)\frac{1}{\nabx^2}\mu_2\right]\,.
\end{align}
This is the Newtonian potential to second order in (Eulerian) perturbation theory. In order to stress the functional dependence we have labeled derivatives w.r.t. $\vc{x}$ by a vertical bar. Similar results for $\Lambda$CDM have been recently obtained in \cite{Rampf:2014mga} with the difference that the transformation studied there is purely spatial, ie the time of the new non-comoving frame is again identified with the proper time of the particles. The authors of \cite{Rampf:2014mga} also treat tensor perturbations which are ignored here.

\section{N-body motions on large scales}

In the previous two sections we presented a set of results describing matter perturbations in $\Lambda$CDM in two commonly used gauges: comoving-synchronous and Poisson gauge. As we will now see, the transformation (\ref{t-xfn}) and (\ref{x-xfn2}) between the two descriptions is very useful in interpreting the the results of N-body simulations on large scales from a General Relativistic perspective.

The metric (\ref{metric-phi}) refers to a comoving coordinate system akin to a Lagrangian description of the dynamics of the particle flow. However, N-body simulations work with a fixed coordinate grid with respect to which the individually labeled particles move. It is therefore natural to ask how the synchronous comoving description can be translated in terms directly comparable to the outcome of an N-body simulation i.e. a set of particle trajectories under the influence of gravity in an Eulerian coordinate system. To connect with General Relativity it is important to first define what is meant by the coordinates used in an N-body simulation in relativistic terms. This is achieved by choosing a correspondence between the coordinates employed - the points of a Euclidean grid $\vc{x}$ and a universally ticking clock $\tau$ - to events in spacetime. In fact, this constitutes a choice of gauge, a choice different from the comoving synchronous gauge. The most natural choice for mapping a simulator's coordinates to spacetime is the Poisson gauge.

In the comoving gauge, the value of $\vc{q}$ labels each worldline and remains constant during particle's motion. It follows that if we express $\vc{x}$ in (\ref{x-xfn2}) in terms of the Poisson gauge time $\tau$, instead of the particle's proper time $t$, we will directly obtain the trajectory of a particle in the Newtonian N-body frame. The coordinates $\vc{q}$ which label the particles in the synchronous frame can then be interpreted as the initial positions in the Newtonian (Poisson) frame: $\vc{q}=\vc{x}_{\rm ini}$. Using (\ref{x-xfn2}) and replacing $t$ by $\tau$ from (\ref{t-xfn}) we thus have for the trajectories of particles to second order
\begin{align}\label{trajectory}
\vc{x}(\tau,\vc{q})_{\rm trajectory}=\vc{q}+\nabla_\vc{q}\Bigg(\frac 3 2 D \Phi  + \frac 9 4 E
    \frac{1}{\fett{\nabla}_{\fett{q}}^2} \mu_2\Bigg) +\left(5 D+\frac{v^2}{a^2}\right)\Bigg(\nabla_\vc{q}\left(V-\Phi^2\right)+\vc{W}\Bigg)\,,
\end{align}
The terms in the first bracket are the Zeldovich approximation and the second order Newtonian result as derived in Newtonian Lagrangian perturbation theory (2LPT) respectively. Counting gradients, they are the dominant terms on subhorizon scales and, as expected, recover the newtonian dynamics. The second bracket is a General Relativistic correction, containing both a scalar and a vector component that are absent in Newtonian theory. Both these terms are second order in the initial potential. Solving the Newtonian equations will reproduce the first bracket, at least for the small initial displacements at early times. The effects of General Relativity on particle motion can then simply be taken into account by adding the second bracket as a correction to the obtained particles trajectories.

It is worth comparing the trajectory (\ref{trajectory}) with the displacement field of \cite{Rampf:2014mga}. The results differ by an extra $v^2$ term in the coefficient of the second bracket which is missing in \cite{Rampf:2014mga}. This can be traced to the fact that the authors of that work do not transform to the poisson gauge but keep the proper time $t$ as their preferred time. The newtonian term (first bracket) is recovered in both approaches but the relativistic second order corrections (dominant on long wavelengths) differ. This seems reasonable and it might be worth investigating which transformation, if any, is better suited to match the coordinates of an N-body simulation.

We now proceed to ask what density field does the above displacement generate? We have already derived the formula for the density perturbation in the synchronous gauge, formula (\ref{density2}). If we simply replace the transformation (\ref{t-xfn}) and (\ref{x-xfn2}) in (\ref{density2}) we then obtain the density in the Poisson gauge. The two expressions correspond to different gauge invariant quantities with different physical meanings \cite{Bardeen:1980kt}. Both would be admissible as a measure for the density contrast but one can note that the poisson gauge density contrast contains a term proportional to the potential $\Phi$ that diverges rapidly in the IR for an inflationary primordial spectrum. This is not a problem on subhorizon scales but is undesirable on superhorizon scales which is our range of interest. We therefore advocate using the synchronous comoving density which is better behaved in the IR
\begin{widetext}
\begin{align}\label{density3}
\Delta(\tau,\vc{x})=&-\frac{3}{2}D\Phi_{|ll}+\frac{9}{8}\left(D^2-E\right)\Phi_{|ll}\Phi_{|mm} +\frac{9}{8}\left(D^2+E\right)\Phi_{|lm}\Phi_{,lm}+\frac{9}{4}D^2\Phi_{|l}\Phi_{|lmm}\nonumber
\\&+10D\Phi\Phi_{|ll }+\frac{15}{4}D\Phi_{|l} \Phi_{|l}\,.
\end{align}
\end{widetext}
Note that we have relabeled $t$ in expression (\ref{density}) with the simulation time $\tau$ without using (\ref{t-xfn}). This makes (\ref{density3}) a gauge invariant quantity in the poisson gauge which reduces to the density in the synchronous gauge. We have however used (\ref{trajectory}) to leading order to pass to an Eulerian description. The first four terms in (\ref{density3}) coincide precisely with the result from (Newtonian Eulerian) 2LPT. Eq. (\ref{density3}) gives precisely the density contrast in the Eulerian gauge \cite{Rampf:2014mga}.

In order to access the importance of the non-Newtonian terms we can compute the power spectrum associated with the density perturbation that is generated by displacing particles according to the displacement field (\ref{trajectory}), ignoring the vector part \cite{Rigopoulos:2013nda}. We compare three cases: Zeldovich approximation alone (ZA), labeled 1 in figure 2, Newtonian 2LPT, labeled 12 in the figure, and the full relativistic result, labeled 1234. We see in the figure that the difference between the full displacement and the ZA is identical to the difference between 2LPT and and the ZA on scales below the horizon, indicating that the relativistic terms are completely negligible. On scales larger than the horizon there is a crossover and the difference between the full result and the Zeldovich approximation is identical to that between the full result and 2LPT, indicating that the 2LPT terms are completely subdominant. The relativistic terms provide a correction to the ZA power spectrum of the order of $1\%$ on scales about 10 times bigger than the horizon at $z=49$. Because the GR correction is at any redshift only apparent on scales larger than the horizon, it is clear that the GR corrections decay once they are within the horizon. For more details see \cite{Rigopoulos:2013nda}.

\begin{figure*}
\includegraphics[width=0.45\textwidth]{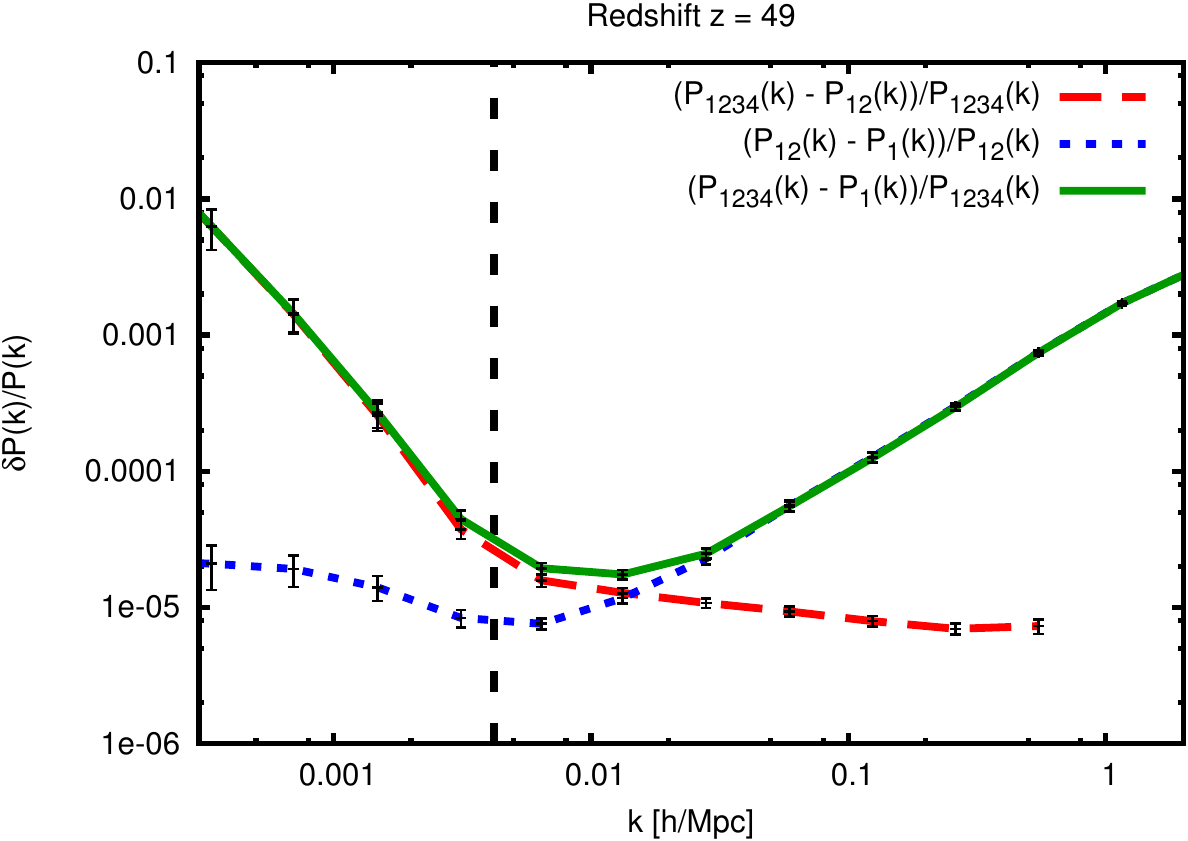}\hspace{0.07\textwidth}
\includegraphics[width=0.45\textwidth]{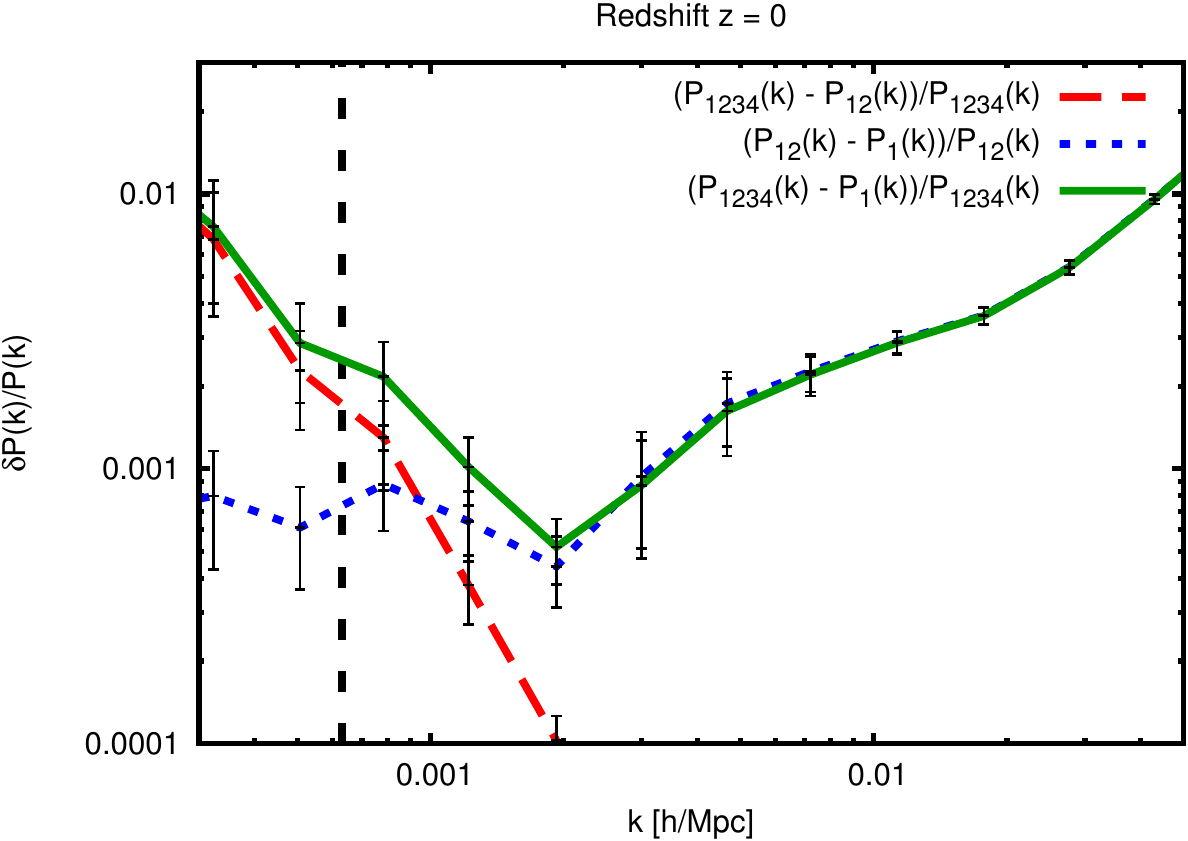}
\caption{The normalized difference in density power spectrum P(k) ({\em left:} at $z=49$, {\em right:} at $z=0$) measures the difference in total variance on a certain scale $k$ from different terms in the displacement field \ref{trajectory}. The solid (green) line shows the deviation of the Zeldovich Approximation (ZA, labeled $\{1\}$) from the full displacement field (labeled $\{1234\}$), the long dashed line (red) shows the deviation of 2nd-order Lagrangian Perturbation Theory (2LPT, labeled $\{12\}$) from the full result, while the short dashed line (blue) shows the deviation of ZA from 2LPT.
We see that on small scales the correction to ZA is entirely due to 2LPT contributions, while on scales larger than the particle horizon (indicated by a vertical dashed line) it is entirely due to the relativistic terms in (\ref{trajectory}), showing the 2LPT terms to be subdominant on such scales. For more details on the simulations behind these plots and the computation of the error bars see \cite{Rigopoulos:2013nda}}\label{fig:pofk}
\end{figure*}

\section{Conclusions}\label{}

In this paper we reviewed a set of results describing CDM perturbations in LCDM cosmology in two commonly used gauges: comoving-synchronous and Poisson gauge. These results, scattered in the literature in one form or another, were presented here in a concise and hopefully clear manner. They were derived from a gradient expansion of the Einstein equations which is fully non-linear on large scales but equivalent to second order perturbation theory for CDM when truncated to second order in $\Phi$. Known newtonian results have been recovered and relativistic corrections identified. We then argued that the gauge transformation between the two gauges can directly provide the long wavelength trajectories of particles in the gauge associated with N-body simulations, the Poisson gauge. The trajectories (\ref{trajectory}) are composed of the Zeldovich approximation, the Newtonian 2LPT result and a relativistic correction which is second order in the metric potential $\Phi$.

It is clear from (\ref{trajectory}) that at linear order the Zeldovich approximation, a Newtonian result, also holds true in the full GR description on all scales. Thus, simulations set up with Zeldovich initial conditions and evolved with the Newtonian dynamical equations produce the correct relativistic particle trajectories to first order in the potential. At second order however here are relativistic corrections which overtake the Newtonian second order result on scales comparable to or exceeding the Horizon. For example, the relativistic corrections to the power spectrum for a simulation box comparable to our current horizon is at the level of $\lesssim 0.01$ for the largest scales contained in the box, independent of redshift as long as the modes are outside the horizon. The impact of this extra power compared to that generated by the ZA initial conditions on precision measurements of the later dynamics is currently unknown. It is worth noting that for simulations started at a relatively a small redshift,  $z\sim 50$, 2LPT initial condition are usually employed. As we have seen, they are in fact subdominant on large scales compared to the relativistic contributions. Although these observations do not necessarily imply that the later subhorizon dynamics are affected in any significant manner, these issues are worth investigating given that large N-body simulations are now used for precision sub-percent measurements of the power spectrum on scales in the BAO regime. We leave this investigation for future work \cite{CRV}.

\section*{Acknowledgements}
C.R. acknowledges the support of the individual fellowship RA 2523/1-1 from the German Research Foundation (DFG). G.R. is supported by the German Research Foundation (DFG) through the TRR33 program 'The Dark Universe'. W.V. is supported by a Veni research grant from the Netherlands Organization for Scientific Research (NWO).


\begin{thebibliography}{99}

\bibitem{Bernardeau:2001qr}
  F.~Bernardeau, S.~Colombi, E.~Gaztanaga and R.~Scoccimarro,
  Phys.\ Rept.\  {\bf 367} (2002) 1
  [astro-ph/0112551].

\bibitem{Malik:2008im}
  K.~A.~Malik and D.~Wands,
  Phys.\ Rept.\  {\bf 475} (2009) 1
  [arXiv:0809.4944 [astro-ph]].

\bibitem{Green:2011wc}
  S.~R.~Green and R.~M.~Wald,
  Phys.\ Rev.\ D {\bf 85} (2012) 063512
  [arXiv:1111.2997 [gr-qc]].

\bibitem{Hwang:2012ra}
  J.~-c.~Hwang and H.~Noh,
  JCAP {\bf 1304} (2013) 035
  [arXiv:1210.0676 [astro-ph.CO]].

\bibitem{Flender:2012nq}
  S.~F.~Flender and D.~J.~Schwarz,
  Phys.\ Rev.\ D {\bf 86} (2012) 063527
  [arXiv:1207.2035 [astro-ph.CO]].

\bibitem{Kopp:2013tqa}
  M.~Kopp, C.~Uhlemann and T.~Haugg,
  JCAP {\bf 1403} (2014) 018
  [arXiv:1312.3638 [astro-ph.CO]].

\bibitem{Villa:2014aja}
  E.~Villa, S.~Matarrese and D.~Maino,
  JCAP {\bf 1406} (2014) 041
  [arXiv:1403.6806 [astro-ph.CO]].

\bibitem{Basset Hlozek}
B. Bassett and R. Hlozek, arXiv:0910.5224 [astro-ph.CO]

\bibitem{Amendola:2012ys}
  L.~Amendola {\it et al.}  [Euclid Theory Working Group Collaboration],
  Living Rev.\ Rel.\  {\bf 16} (2013) 6
  [arXiv:1206.1225 [astro-ph.CO]].


\bibitem{Bernardeau:2013oda}
  F.~Bernardeau,
  arXiv:1311.2724 [astro-ph.CO].

\bibitem{Rasera:2013xfa}
  Y.~Rasera, P.~-S.~Corasaniti, J.~-M.~Alimi, V.~Bouillot, V.~Reverdy and I.~Balmès,
  Mon.\ Not.\ Roy.\ Astron.\ Soc.\  {\bf 440} (2014) 1420
  [arXiv:1311.5662 [astro-ph.CO]].

\bibitem{Rigopoulos:2012xj}
  G.~Rigopoulos and W.~Valkenburg,
  Phys.\ Rev.\ D {\bf 86} (2012) 043523
  [arXiv:1203.2796 [astro-ph.CO]].

\bibitem{Rampf:2012pu}
  C.~Rampf and G.~Rigopoulos,
  Mon.\ Not.\ Roy.\ Astron.\ Soc.\ Lett.\  {\bf 430} (2013) L54
  [arXiv:1210.5446 [astro-ph.CO]].

\bibitem{Rampf:2013ewa}
  C.~Rampf and G.~Rigopoulos,
  Phys.\ Rev.\ D {\bf 87} (2013) 12,  123525
  [arXiv:1305.0010 [astro-ph.CO]].

\bibitem{Rigopoulos:2013nda}
  G.~Rigopoulos and W.~Valkenburg,
  arXiv:1308.0057 [astro-ph.CO].

\bibitem{Rampf:2013dxa}
  C.~Rampf,
  Phys.\ Rev.\ D {\bf 89} (2014) 063509
  [arXiv:1307.1725 [astro-ph.CO]].

\bibitem{Comer:1994np}
  G.~L.~Comer, N.~Deruelle, D.~Langlois and J.~Parry,
  Phys.\ Rev.\ D {\bf 49} (1994) 2759.

\bibitem{Stewart:1994wq}
  J.~M.~Stewart, D.~S.~Salopek and K.~M.~Croudace,
  Mon.\ Not.\ Roy.\ Astron.\ Soc.\  {\bf 271} (1994) 1005
  [astro-ph/9403053].

\bibitem{Adamek:2014xba}
  J.~Adamek, R.~Durrer and M.~Kunz,
  arXiv:1408.3352 [astro-ph.CO].

\bibitem{Russ:1995eu}
  H.~Russ, M.~Morita, M.~Kasai and G.~Borner,
  Phys.\ Rev.\ D {\bf 53} (1996) 6881
  [astro-ph/9512071].

\bibitem{Matarrese:1997ay}
  S.~Matarrese, S.~Mollerach and M.~Bruni,
  Phys.\ Rev.\ D {\bf 58} (1998) 043504
  [astro-ph/9707278].

\bibitem{Rampf:2014mga}
  C.~Rampf and A.~Wiegand,
  arXiv:1409.2688 [gr-qc].

\bibitem{Bardeen:1980kt}
  J.~M.~Bardeen,
  Phys.\ Rev.\ D {\bf 22} (1980) 1882.

\bibitem{CRV}
S.~Casas, G.~Rigopoulos and W. Valkenburg, in progress


\end{thebibliography}
\end{document}